\documentclass[12pt]{article}
\usepackage[utf8]{inputenc}
\usepackage[a4paper, margin=1in,left=2.5cm, right=2.5cm, top=2.5cm]{geometry}

\usepackage{amsmath}
\usepackage{float}
\usepackage{graphicx}
\usepackage{subcaption}
\usepackage{wrapfig}
\usepackage{mathptmx} 
\usepackage{helvet}

\title{\huge{\textbf{Investigation of atomic and molecular rates in plasma-edge simulations through experiment-simulation comparisons}}}

\author{
        A. C. Williams \\
                Department of Physics\\
        University of York\\
        Heslington, York, UK}
\date{\today}

\begin{document}
\maketitle
\thispagestyle{empty}

\clearpage 
\begin{abstract}
Divertor plasma detachment is likely needed for the function of magnetically confined nuclear fusion. It greatly reduces the particle and heat flux incident on a target, and thus reduces the sputtering and heat loading on the target. It is therefore advantageous to have accurate simulations of plasma behaviour in the divertor to design future Tokamak divertors.\\ A common simulation package used for this purpose is SOLPS-ITER, which has previously been observed to be underestimating the contribution of molecular effects to plasma detachment. To correct this, its reaction rate for molecular charge exchange was altered. This alteration resulted in a significant increase in the density of $D_2^+$ ions within the divertor. The particle balance and $D\alpha$ emission data was then computed from the simulation outputs. These were compared to experimental data and a set of post-processing routines for the original simulations that predicts the effect of a corrected charge exchange rate.
\\ In comparison to the post-processing predictions: the increase in Molecular activated Recombination and the depth of the target flux rollover was less than predicted. This is due to the post-processing not being self consistent and it does not account for how the alterations in reaction rate would affect the background plasma. Compared to the experimental data: the correction showed a significant improvement in the agreement of general trends, such as the increased MAR reaction rate and the rollover of target flux. However, there were still significant disagreements in the magnitudes of various reaction rates, suggesting further improvements to SOLPS-ITER are still needed.

\end{abstract}
\thispagestyle{empty}

\clearpage
\pagenumbering{arabic}
\maketitle

\section{Introduction}

Developing a Nuclear fusion power plant is important due to its high utility as  a  sustainable  source  of  electricity. Fossil fuels are not sustainable for long term use and humanity's power consumption only increases over time \cite{Energy consumption}. Many renewable sources have a low energy density, requiring significant landmass \cite{electircity generation density}. Due to varying output, they also require large amounts of energy storage in specialised geographies \cite{Gravitational energy storage} and thus many city areas not near such locations could easily be cut off from power during natural disasters. Nuclear fusion provides an option for mass electricity production that is safe \cite{Nuclear fusion safety}, has large supplies of fuel \cite{Nuclear fusion fuel}, can be built close to city areas \cite{Nuclear fusion safety} and has reliable, variable power production.\\\\
Nuclear fusion is the act of colliding two nuclei, at a high enough energy to overcome the electromagnetic force, so they can react and fuse into a heavier nucleus. The energy released during this reaction is determined by the difference in binding  energy per nucleon for each of the nuclei. This directly converts the mass of the atoms into energy \cite{Energy and mass equation} . This released energy can then be captured to produced electricity. \\\\
The most developed design for a nuclear fusion power plant is the Tokamak. A Tokamak works by heating hydrogenic gas into a plasma. This plasma is then confined using a set of toroidal field to create a torus shape, which then has a plasma current pushed through it causing it leading to a helical magnetic field structure, confining the plasma \cite{Tokamaks DT Reaction rates Tokamak and plasma shape and divertor}. This electromagnetic confinement allows for a high enough plasma density and large enough confinement time for the fusion collisions between hydrogen atoms to occur. However, there is a limit to the cross-field transport that can occur due to the helical field structure and so some ions will escape confinement. To prevent damage to the Tokamak components, this escaping plasma is diverted into a structure designed to absorb the large heat flux, called the divertor \cite{Tokamaks DT Reaction rates Tokamak and plasma shape and divertor}. \\\\
Our experiment will focus on the divertor in a Tokamak, called Tokamak à configuration variable (TCV), that uses Deuterium ($H^2_1$) as fuel.
This is because a deuterium-deuterium plasma comes closest to the ideal deuterium-tritium mix without having to produce and handle radioactive tritium.
All currently existing Tokamaks are experimental in nature. In order to create a working Nuclear fusion power plant, there are several problems with Tokamaks that need to be addressed. One of these issues is heat dissipation within the divertor \cite{Magnetic topologys}. The Current material limitations of the divertor are that they can withstand heat-fluxes of approximately 10 $MW/m^2$ \cite{10MW limit directly} \cite{10MW limit for DEMO} \cite{also 10MW limit directly}.
Future designs for power plants, such as DEMO, are estimated to have heat fluxes in excess of 50 $MW/m^2$ \cite{DEMO design concepts} \cite{Over 50MW directly} and so this heat must be dissipated before reaching the divertor target.\\\

This heat flux can be greatly reduced by inducing detachment in the divertor. Detachment happens when particle, power and momentum losses simultaneously \cite{Plasma detachment from divertors and limiters} occur as the temperature reduces at the target, this then results in a reduction of the target particle flux and the surface recombination heat load that it brings with it \cite{potential detachment definition reference}. As a result, a good understanding of how detachment works is vital for heat dissipation and the longevity of future machines. \\
Due to the extremely high cost of construction and running Tokamaks, the ability to understand, model and then simulate divertor detachment and its processes is vital in the understanding of detachment and the design and operation of future Tokamaks. As a result, a long running history of Tokamak simulation codes have been produced, the most widely used of which is the SOLPS-ITER suite which models the edge region of the plasma and was built to develop the ITER divertor.

SOLPS-ITER is a combination of parallelized EIRENE Monte Carlo neutrals solver code with the B2.5 fluid plasma solver \cite{parallelized EIRENE code with the B2.5 fluid plasma solver} which together simulate divertor detachment and its processes.\\\\
One key set of reactions are the molecular interactions within the plasma. The molecules ($D_2$) undergo electron impacts which vibrationally excite the molecules ($D_2 (\nu)$). These can then undergo various reaction paths to create molecular ions ($D_2^+$) and these ions subsequently react with the plasma to result in power and particle losses. Of particular importance to this project is the molecular activated recombination (MAR) reaction of $D_2^+$. The $D_2^+$ ion is formed by molecular charge exchange with a $D^+$ ion, it then recombines with an electron to form $D_2 (\nu) + D$ where the excited $D_2 (\nu)$ atoms then decays releasing a photon. This results in particle (ion) losses within the plasma and radiating power thus inducing power losses. These photons can then be detected using spectrometers observing the divertor. One experimental technique known as BaSPMI uses Balmer line emissions to infer the contributions of the various plasma-atom/molecule interactions to the hydrogen emissions. From these contributions it then infers the  experimental values for the rates of different reactions within the divertor \cite{Detachment definition also reactions also on molecule chains and dissassiation at low temp everything created via charge exchange}, which the simulations can then be compared against. \\\\  
My project is based on altering the SOLPS-ITER reaction rates to help better model the divertor behaviour and this formation of detachment. The current model of SOLPS predicts a low $D_2^+$/$D_2$ ratio for TCV when the electron temperature is less than 5ev \cite{EIRENE simulations predicting D_2 as negligible}. As a result of this, simulations predict that the number of molecular reactions involving $D_2^+$, and thus their predicted impact on the formation of detachment, is low. However, this disagrees with experimental data that shows large numbers of molecular reactions involving $D_2^+$ occurring during the onset of detachment \cite{Main paper on MAR EIR and ionisation rates}.
It is hypothesised that the low $D_2^+$/$D_2$ ratio is caused by an incorrect mass re-scaling for the hydrogen molecular charge exchange rate to deuterium. This lower reaction rate then results in the reduced creation of $D_2^+$ molecules and thus a lower $D_2^+$ density. In this project we will alter the reaction rate for for Deuterium charge exchange to remove this re-scaling \cite{The reaction rates we're using when altered the mass re-scaling}.\\ 
We will then observe how this changes the density of $D_2^+$ ions, the contribution of molecular reactions involving $D_2^+$ to the onset of detachment and if these simulations align with experimental data. From these simulations, it is hoped we can develop a better understanding of the role molecular reactions, particularly with $D_2^+$, play within detachment. From this, a more accurate model of detachment could be developed.\\\\

\section{Theory}

Detachment is caused by a variety of atomic and molecular reactions which induce simultaneous power, momentum, and particle losses \cite{Detachment definition} \cite{Detachment definition also reactions also on molecule chains and dissassiation at low temp everything created via charge exchange}. A key feature of detachment is the reduction of the ion target flux, which can be reduced by lowering the ion source or recombining ions volumetrically into neutrals
This greatly decreases damage to the divertor by limiting sputtering on the target. As a result, monitoring the particle balance of ions sources and sinks is important for the longevity of the divertor.\\

The main reactions of importance to this paper within the divertor are either atomic or molecular and are as follows \cite{Molecular equations reference} \cite{Equations for deuterium}:\\

\begin{center}
\begin{tabular}{|c|c|c|} 
 \hline
Atomic Ionisation & $D + e^- -> D^+ + 2e^-$ & [1] \\ [0.5ex] 
 \hline
  Molecular Charge Exchange & $D^+ + D_2 -> D + D_2^+$ & [2] \\
 \hline
 Radiative Electron Ion Recombination (EIR) & $D^+ + e^- -> D(\nu)$ & [3]  \\ 
 \hline
 Non-radiative Electron Ion Recombination (EIR) & $D^+ + e^- + e^- -> D + e^-$ & [4]  \\ 
 \hline
 Dissociative Recombination & $D_2^+ + e^- -> D_2 (\nu) + D$ & [5] \\
 \hline
 Dissociative excitation & $D_2^+ + e^- -> D_2 (\nu) +D^+ + e^-$ & [6]  \\
 \hline

 Molecular Ionisation & $e^- + D_2 -> D_2^+ + 2e^-$ & [7] \\ [1ex] 
 \hline
\end{tabular}
\end{center}

Molecular activated recombination acts in a chain of reactions forming $D_2^+$ via reaction 1 and then recombining it via reaction 5 to remove ions from the plasma. Ignoring transport of molecular ions, which is justified as molecular ions have a short lifetime, the $D_2^+$ ion density is determined by the ratio between the sum of the reactions that create $D_2^+$ divided by the sum of reactions that destroy $D_2^+$. At lower temperatures of around 5ev, the Molecular Ionisation rate decreases significantly \cite{Detachment definition also reactions also on molecule chains and dissassiation at low temp everything created via charge exchange}. This means that, at low temperatures, the $D_2^+$ ion density is overwhelmingly determined by the rate of Molecular charge exchange and thus, Molecular activated recombination is also dependent of the rate of Molecular charge exchange.\\

Electron ion recombination (EIR) 
is a direct ion sink of abundant $D^+$ ions and so has often been seen as an important part of detaching the plasma. However, recent studies have found that, for TCV, the rate of EIR has minimal impact on the initial formation of detachment \cite {Role of particle sinks and sources (EIR not important)}.
A more significant effect in the onset of detachment is that of MAR and power starvation \cite{understanding atomic processes (EIR not important power starvation important)}. As the temperature of the plasma decreases, each neutral atom has a lower average energy and so requires more energy from collisions to ionise. Thus, at lower temperatures, obtaining the same amount of ionisation will cost more power and require a higher neutral density. This reduces or flattens the ion source leading to a decrease or flattening of the ion source. \\

The experimental counterpart to the simulation is performed using the spectroscopic technique called BaSPMI that analyses the Balmer line emissions from the divertor. The plasma-atom interactions are measured using an altered version of the technique described in \cite{understanding atomic processes (EIR not important power starvation important)} by observing medium n Balmer lines (n = 5,6,7). These Balmer lines are then analysed using a variety of techniques to determine the contributions, and thus the rates of, atomic ionisation and recombination, after compensating for the molecular contributions to the medium n Balmer lines \cite{Detachment definition also reactions also on molecule chains and dissassiation at low temp everything created via charge exchange}.\\ The plasma-molecular interactions are calculated using the $D\alpha$ and $D\beta$ lines. The $D_2^+$ and $D^-$ contributions are separated using the ratio between the $D\alpha$ and $D\beta$ lines. Determining the rates of MAR and MAI is more complicated from here,  as if a reaction counts as MAR or MAI can depend on if the chain of reaction begins using charge exchange or molecular ionisation, as a result the MAR/MAI ratio is simply calculated using fixed reaction rates taken from AMJUEL \cite{Detachment definition also reactions also on molecule chains and dissassiation at low temp everything created via charge exchange}.\\

Recent studies have observed the difference between experimental and simulation data can be explained by the influence of MAR reactions as an ion sink \cite{plasma-molecule on particle balance (MAR important)}. This is surprising as current SOLPS simulations predict a very low $D_2^+$ density in the plasma and as a result a low rate of MAR \cite{EIRENE simulations predicting D_2 as negligible}, contradicting the experimental data. Therefore, a post-processing routine was created, which can be applied to simulation results, to approximate the adjusted reaction rates at higher $D_2^+$. However, this post-processing does not account for the the way these increases in reaction rates may affect the rest of the plasma and so a self consistent adjustment to the SOLPS simulation is needed \cite{Main paper on MAR EIR and ionisation rates}.\\ 

This difference in the measured and simulated $D_2^+$ density is theorized to be due to an inaccuracy in the calculation of the deuterium molecular charge exchange rate in EIRENE simulations. The reaction rate for charge exchange is based on two factors. One is the the distribution of vibrational states for the $D_2$ molecules. This vibrational state distribution is due to electrons impacting on the molecules and is thus dependent on the electron temperature ($T_e$). The different vibrational states have different reaction cross sections $<\sigma V>$ for the reaction of $p + H_2 = H + H_2^+$ and thus affect the reaction rate. The second is the ion temperature ($T_i$) as the reaction is due to the collision between a $D^+$ ion and the $D_2$ molecule and thus the relative velocity of the two particles is a key factor. SOLPS calculates the total reaction rate for $D_2$ by taking the rate of charge exchange for $H_2$ and altering them for Deuterium's increased mass \cite{Correct vibrational states graph? go over for exact names}. A correct method of doing this would solely re-scale the $T_i$ dependence, as Deuterium has double the mass of Hydrogen and thus would have half the velocity at the same temperature. It would also require taking into account the chemical differences between Deuterium and Hydrogen in terms of their differences in vibrational states \cite{The reaction rates we're using when altered the mass re-scaling}.\\

The hypothesis of this experiment is that the ERIENE code simplifies this calculation by taking the full calculation of the reaction rate with respect to temperature, assumes $T_e = T_i = T$, and then halves the temperature for the entire reaction rate \cite{EIRENE code user manual}. This means that the electron temperature dependence is also re-scaled and this is believed to be unjustified. This then results in the charge exchange rate, at the temperatures relevant to detachment, being underestimated by 1 - 2 orders of magnitude. As a result, the amount of $D_2^+$ ions produced is greatly decreased and thus reducing the amount of MAR can occur. Thus, we will investigate the effects of correcting the reaction rate for molecular charge exchange for deuterium. Kukushkin attempted a more accurate calculation of the $D_2$ reaction rate by two methods. One by removing the temperature re-scaling and treating $D_2$ the same as $H_2$ and another by a re-scaling of the $T_i$ rate alone \cite{Molecular equations reference}. His results found a minimal difference between the two approaches, suggesting the main dependence of the reaction rate was on $T_e$. However, his approach neglected the chemical differences between Hydrogen and Deuterium and of their vibrational states and energy distributions. Thus, Reiter performed a more in depth analysis which took these chemical differences into account, assuming identiical vibrational distributions between hydrogen and deuterium. This concluded that the $H_2^+$ charge exchange rate is within 95\% accuracy of the correct reaction rate for $D_2$ molecules \cite{The reaction rates we're using when altered the mass re-scaling}. As such, this justifies our approach of enforcing the hydrogen molecular charge exchange rate while simulating a deuterium plasma.

\section{Method}

\begin{figure}[!h]
     \centering
     \begin{subfigure}[b]{0.45\textwidth}
         \centering
         \includegraphics[width=.75\linewidth]{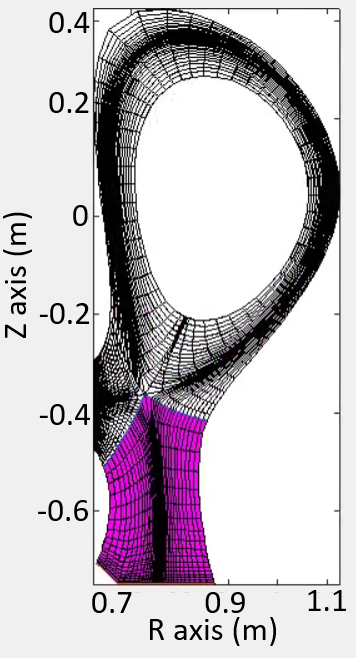}
         \captionof{figure}{}
         \label{fig:Tokamak Grid}
     \end{subfigure}
     \hfill
     \begin{subfigure}[b]{0.45\textwidth}
         \centering
         \includegraphics[width=.75\linewidth]{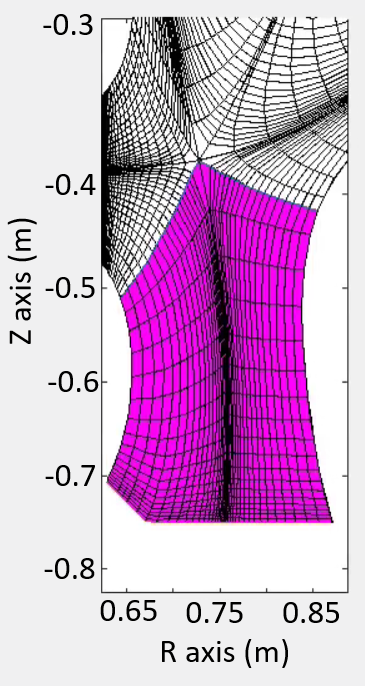}
         \captionof{figure}{}
         \label{fig:Tokamak Divertor Grid}
     \end{subfigure}
     \hfill
        \caption{The CARRE meshgrid used in the experiment, with the divertor region highlighted in purple and seen in greater detail in figure \ref{fig:Tokamak Divertor Grid}}
        \label{fig:Carre meshgrid}
\end{figure}

The simulations were performed using the SOLPS-ITER simulation package. This package consists of the grid generator CARRE, the fluid plasma code B2.5 and the neutral particle transport code EIRENE along with various other additional modules  \cite{SOLPS-ITER Dashboard}. The code assumes toroidal symmetry and so is calculated in 2D, with the x axis being parallel to the poloidal direction and the y axis being radial (perpendicular to the flux surfaces). The grid generator CARRE creates a grid mesh of cells in the geometry of a Tokamak, with greater cell density placed towards the target due to the increased gradients occurring there. An image of the numerical grid used is included in figure (\ref{fig:Tokamak Grid}) with the area in purple being the area that will be analysed and referred to as the outer divertor, seen more clearly in figure (\ref{fig:Tokamak Divertor Grid}).\\ 

This grid is then used as an input for Eiriene and B2.5. EIRENE uses a linearized Boltzmann transport equation.
\begin{equation}
\label{Boltzmann transport equation}
\begin{aligned}
{\left[\frac{\partial}{\partial t}+\mathbf{v} \cdot \nabla_{\mathbf{r}}+\frac{\mathbf{F}(\mathbf{r}, \mathbf{v}, t)}{m} \cdot \nabla_{\mathbf{v}}\right] f(\mathbf{r}, \mathbf{v}, t)+\Sigma_{t}(\mathbf{r}, \mathbf{v})|\mathbf{v}| f(\mathbf{v}) } &=\int d^{3} v^{\prime} C\left(\mathbf{v}^{\prime} \rightarrow \mathbf{v}\right)\left|\mathbf{v}^{\prime}\right| f\left(\mathbf{v}^{\prime}\right) \\
&+Q(\mathbf{r}, \mathbf{v}, t)
\end{aligned}
\end{equation}

With the variables, Position vector r, pre-collision velocity $v^{\prime}$, post-collision  velocity vector v, time t, a single particle distribution function f(), Any external particle source Q and the collision kernel (the redistribution function) C \cite{EIRENE code user manual}. This equation is solved using a Monte Carlo method. It takes the initial particle source Q and probability for redistribution, calculated using the motion of the particles between collisions T and the collision kernel C. A random distribution values is selected for Q, T and C and a Markoff Chain is created with Q as the initial distribution. This chain then produces a set of responses R which are then sampled and cut down to a satisfactory confidence interval using their standard deviation. These responses are then used to calculate the transport of neutrals across each grid cell \cite{EIRENE code user manual}. Due to this method of calculation, there is always Monte Carlo noise in the resulting simulations.\\

The B2.5 code simulation of the fluid plasma uses a 2d slice of of the mesh grid and assumes toroidal symmetry. It uses the mean field approximation and so treats all the particles inside of a cell as a one body system \cite{David's thesis} and then solves conservation equations for the charge, momentum, density and thermal energy for each cell. Of key note for this paper is that it assumes $T_e = T_i - T$ and $n_e = n_i = n$ when interrogating the reaction rates.\\\\

In terms of error, we are basing these simulations on simulations already validated by Alex Fil and Kevin Verhaegh which have been used in peer reviewed articles \cite{Detachment definition} \cite{Main paper on MAR EIR and ionisation rates} \cite{Alex's paper}. A key measurement of accuracy is if the simulations, running on more modern version of SOLPS-ITER, correspond to the original simulations run by Alex - see appendix A for futher detail. Alex's simulations were based on the TCV shot 52065 which was studied experimentally. Once the correction had been performed, the results of the simulations were validated via sensitivity scans of the input parameters. By varying the input parameters of the simulation, we can measured how stable the simulations are and observed the results to consistent and stable - see appendix B for further details. \\\\

Key parameters include the fueling rate for the various simulations, which were originally set to 0.6, 1, 1.3, 2 and 2.4 $\times 10^{21}$ nuclei for Alex's simulations. These were then expanded to include 0.7, 0.8 and 0.9 $\times 10{21}$ nuclei to better observe in initial increase in MAR rates. For the original rates, an additional simulation at 2.7 $\times 10{21}$ nuclei fueling rate was performed to achieved the same electron density at the separatrix as the corrected rates at 2.4 $\times 10{21}$ nuclei fueling rate, due to the electron density being used as the control parameter. The input power was 400 KW split evenly between electrons and ions. The pumping model used treated the whole wall as a pumping surface with 1\% of impacting ions and neutrals being removed. The carbon sputtering rate was set at a constant sputtering rate where 3.5\% of impacts from neutrals and ions caused carbon sputtering. The main variable change was the reaction rate of deuterium molecular charge exchange vs electron temperature which was altered to the same as the hydrogen's, as it is within 5\% of the reaction rate obtained by Reiter \cite{The reaction rates we're using when altered the mass re-scaling}. As the SOLPS program halves the temperature for the inputted reaction rate, this is corrected by doubling the temperature for the corrected input reaction rate in the AMJUEL file thus, upon SOLPS halving, it returns to the correct reaction rate.\\
 
SOLPS-ITER runs by solving a it's differential equations iteratively from its starting conditions and so its simulations must be converged.  Convergence was determined by observing if the key values ($T_e$,$T_i$,$n_e$, plasma power, particle balance and particle radiation) remained consistent over 0.05 seconds in timesteps.\\

Many key values are outputted directly by SOLPS and so can be extracted and plotted directly ($T_e$, species densities, etc). However, some values are not outputted by SOLPS and must be calculated such as the molecular reaction rates. As these reactions work in chains, the overall effects of the reaction depends on a combination of reactions together. These combinations aren't recorded directly and so are instead calculated using fixed rates, whereas the simulation contains Monte Carlo noise and so there will errors in the calculation of molecular reactions, which were minimised by reduced Monte Carlo noise as much as possible via averaging over time steps but not removed.

\section{Results}\label{results}
\begin{figure} [!ht]
\centering
  \centering
  \includegraphics[width=.65\linewidth]{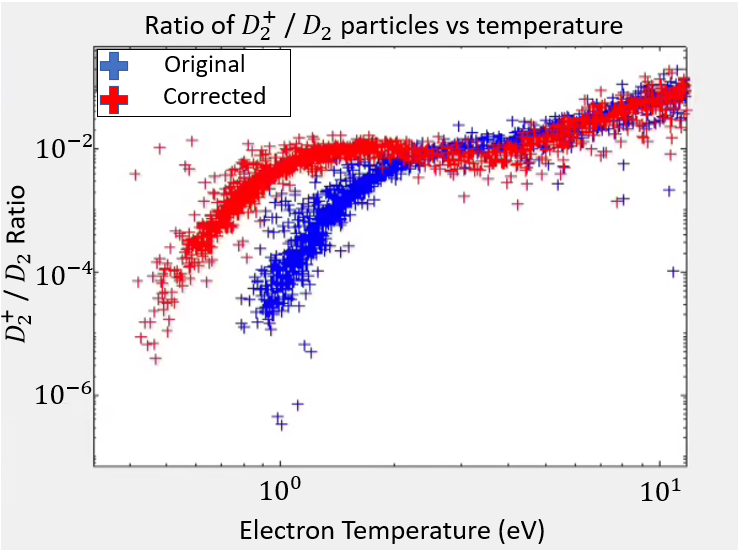}
  \captionof{figure}{A plot comparing the density of $D_2^+$ ions for the original and correct reaction rates. The data shows a significant increase of $D_2^+$ density for the corrected rate at low temperatures. The difference in density is around 2 orders of magnitude when below 1ev.}
  \label{fig:D2+ density ratio}
\end{figure}

Observing the effect of changing the reaction rate, we see that this change has significantly altered the density of of $D_2^+$ ions within the divertor. Looking at figure \ref{fig:D2+ density ratio} we can see a logarithmic scatter plot of  the $D_2^+$/$D_2$ ratio for each SOLPS-ITER grid cell for the original and corrected reaction rates. Above 2ev there is minimal changes, but once we get to below 1ev there is a significant difference in the densities. It can be observed that at 1ev, the increased rate of charge exchange results in around a $10^2$ difference between corrected and original $D_2^+$/$D_2$ and this difference continues to lower temperatures. This suggests the corrected reaction rate has resulted in a significant increase in charge exchange at lower temperatures and this should have a knock on effect on reactions which require $D_2^+$ ions.\\

\begin{figure}[!ht]
    \centering
    \begin{subfigure}[b]{0.55\textwidth}
        \centering
        \includegraphics[width=.95\linewidth]{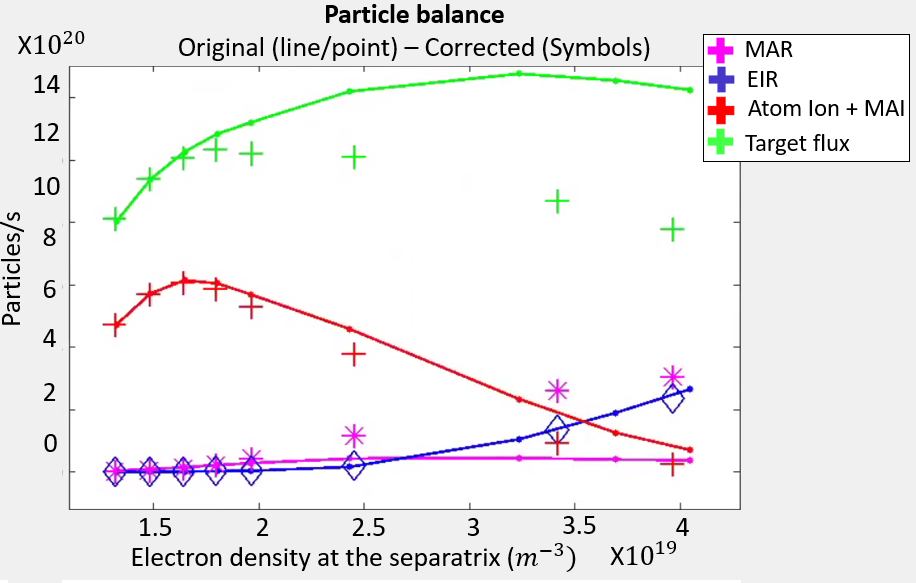}
        \captionof{figure}{}
        \label{fig:Original vs corrected particle balance}
    \end{subfigure}
    \hfill
    \begin{subfigure}[b]{0.44\textwidth}
        \centering
        \includegraphics[width=.95\linewidth]{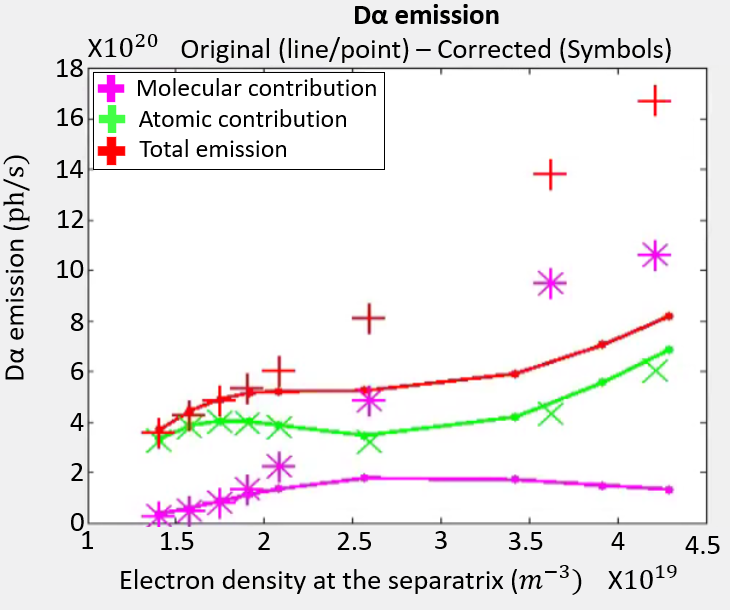}
        \captionof{figure}{}
        \label{fig:Original vs corrected D alpha emission}
    \end{subfigure}
    \hfill
    \caption{Plot \ref{fig:Original vs corrected particle balance} compares the ion sources and sinks to the particle balance as a function of the upstream electron densities for the original and corrected D2+ density. The data shows a significant increase in MAR and a slight decrease in ionisation. This leads to a lower particle flux at the target, reducing the number of ions incident on the target, indicating detachment. Plot \ref{fig:Original vs corrected D alpha emission} compares the $D \alpha$ contribution emissions as a function of the upstream electron densities for the original and corrected molecular charge exchange rates. The data shows a significant increase in Molecular contribution, leading to an increased brightness in the $D \alpha$ emission overall.}
    \label{fig:Original}
\end{figure}  
We can observe the particle balance and the reactions that contribute to it along with the corresponding produced $D\alpha$ emissions in figure \ref{fig:Original}, which has been integrated over the outer divertor leg. In \ref{fig:Original vs corrected particle balance} we can see that the majority of the reactions remain unchanged by the increase in $D_2^+$ ions, however we can also see that initially the rate of MAR remains unchanged but it significantly increases as the density ramping increases. This increase corresponds with the temperature decreasing and thus entering the region where we now have the significantly increased $D_2^+$ ion density.
As a result, we see a significant increase in the contribution of MAR to the ion sinks and correspondingly we observe an earlier and deeper detachment in the particle flux reaching the target. Interestingly, we also note a small decrease in the ion sources due to ionisation, which is mainly due to a reduction in the atomic ionisation. 
We can also see in figure \ref{fig:Original vs corrected D alpha emission} that the molecular contribution to the $D\alpha$ increases significantly and becomes dominant over the atomic contribution as seen in paper \cite{Detachment definition also reactions also on molecule chains and dissassiation at low temp everything created via charge exchange}.\\

\begin{figure}[!ht]
    \centering
    \begin{subfigure}[b]{0.51\textwidth}
        \centering
        \includegraphics[width=.95\linewidth]{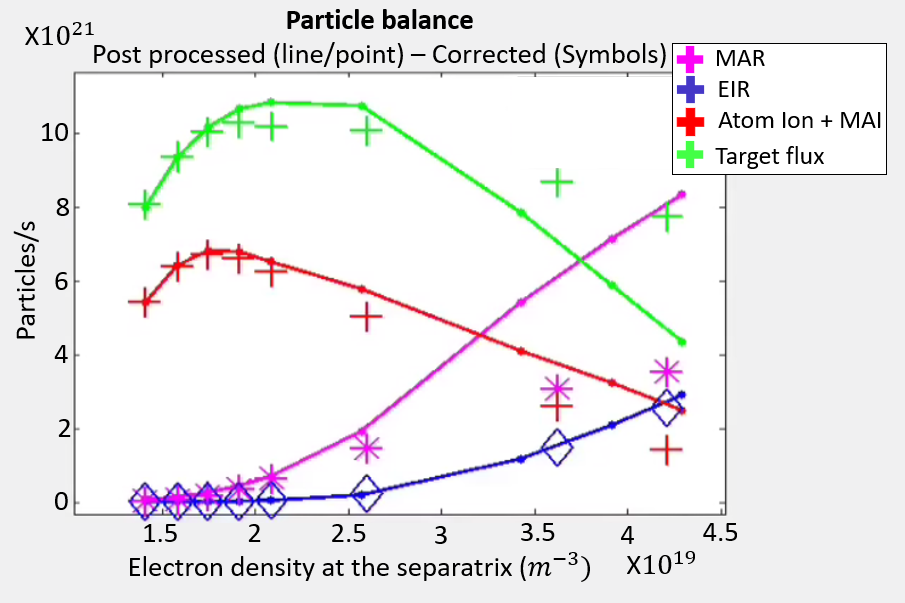}
        \captionof{figure}{}
        \label{fig:Post processed vs corrected particle balance}
    \end{subfigure}
    \hfill
    \begin{subfigure}[b]{0.46\textwidth}
        \centering
        \includegraphics[width=.95\linewidth]{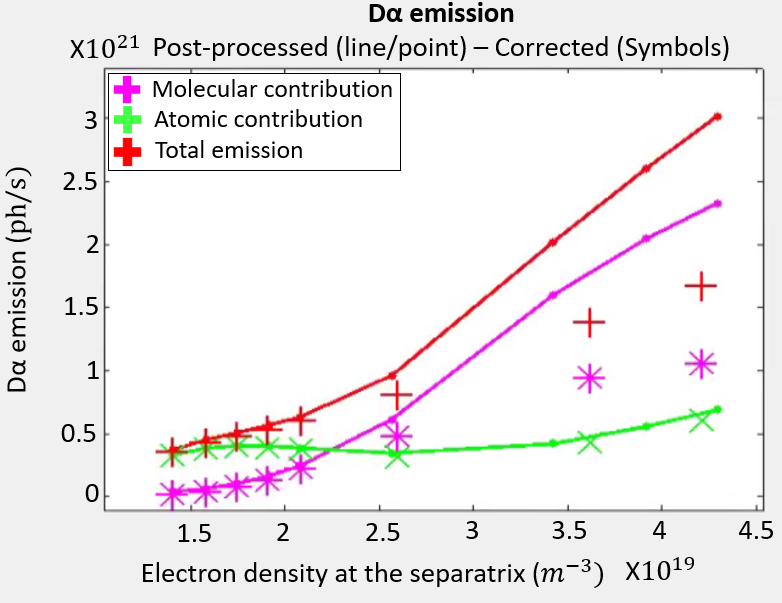}
        \captionof{figure}{}
        \label{fig:Post processed vs corrected D alpha emission}
    \end{subfigure}
    \hfill
    \caption{Plot is \ref{fig:Post processed vs corrected particle balance} comparing the post-processing predictions, for ion sources and sinks contributing to the particle balance as a function of the upstream electron densities, to those of the corrected simulations. The data shows a lower than expected increase in MAR for the corrected data and as a result a higher target particle flux and lesser detachment. Plot \ref{fig:Post processed vs corrected D alpha emission} comparing the post-processing prediction for the $D\alpha$ contribution emissions as a function of the upstream electron densities, to the corrected simulations. The data shows a lower than expected amount Molecular contribution, leading to an decreased brightness in the $D\alpha$ emission overall.}
    \label{fig:Post-processed}
\end{figure}

We can compare these simulation results to the post processing predictions created from the routines made in \cite{Main paper on MAR EIR and ionisation rates}. This can be seen in figure \ref{fig:Post processed vs corrected particle balance}, where we observe a significantly lower MAR rate in the corrected rate compared to the post processing prediction. As a result of this, we can see a far greater reduction of the target particle flux in the post-processing predictions and thus a significantly deeper detachment. We also see a corresponding increase in the molecular contribution to $D\alpha$ emission and a larger $D\alpha$ emission over all. A point of note is that the decrease in ionisation is not predicted by the post-processing and remains the same as the original simulations atomic ionisation rate.\\

\begin{figure}[!ht]
     \centering
         \centering
         \includegraphics[width=.99\linewidth]{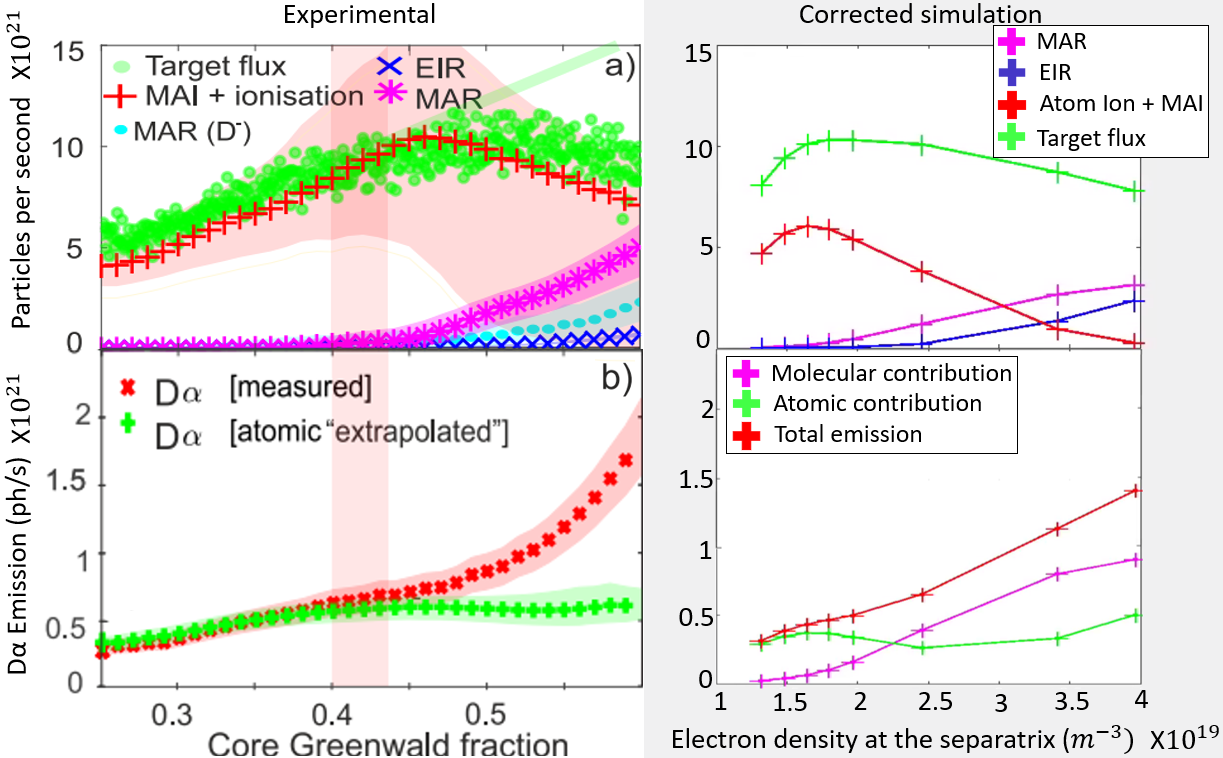}
        \caption{A comparison of the corrected simulations to BaSPMI experimental data for shot number  56567 (a repeat of 52065). The two sets of data range differently on the x axis, covering approximately 0.4 - 0.55 of the Greenwald fraction. While there is an agreement of the general trends between the two data sets, the exact magnitudes of the reactions are significantly different.}
        \label{fig:Experimental data}
\end{figure}

Finally, we can compare the simulation data to experimental data taken from TCV. The experimental data is taken from $D\alpha$ measurements of shot number 56567 (a repeat of shot 52065) which were then processed using BaSPMI analysis and published in \cite{Main paper on MAR EIR and ionisation rates}. A copy of these published results is compared to the simulations in figure \ref{fig:Experimental data}. The experimental data uses the Core Greenwald fraction as it's control as opposed to the electron density, but we can say the lowest simulation fueling density corresponds to approximately the point of rollover for the experimental data. We can observe a general agreement in the trends of the two data sets for the reaction rates of the contributions to the particle balance and in the slight rollover of the target particle flux. We can also see the domination of molecular contributions to the $D\alpha$ emission, although this is of a greater magnitude in the experimental case. \\\\

\section{Discussion} \label{discussion}
\begin{figure} [!ht]
\centering
  \centering
  \includegraphics[width=.8\linewidth]{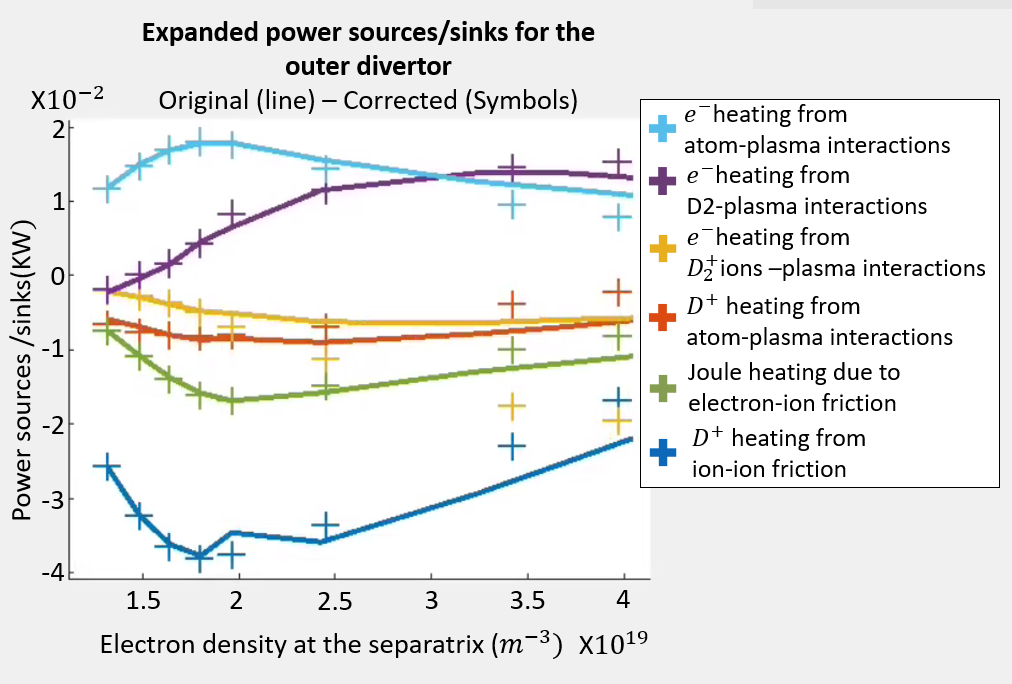}
  \captionof{figure}{The data shows the changes in power balance between the corrected and original rates vs the electron density at the seperatrix. We can see minimal changes in the momentum balance between the two cases, outside of $D_2^+$ ion plasma interactions) }
  \label{fig:Power balance total}
\end{figure}

\begin{figure} [!ht]
\centering
  \centering
  \includegraphics[width=.65\linewidth]{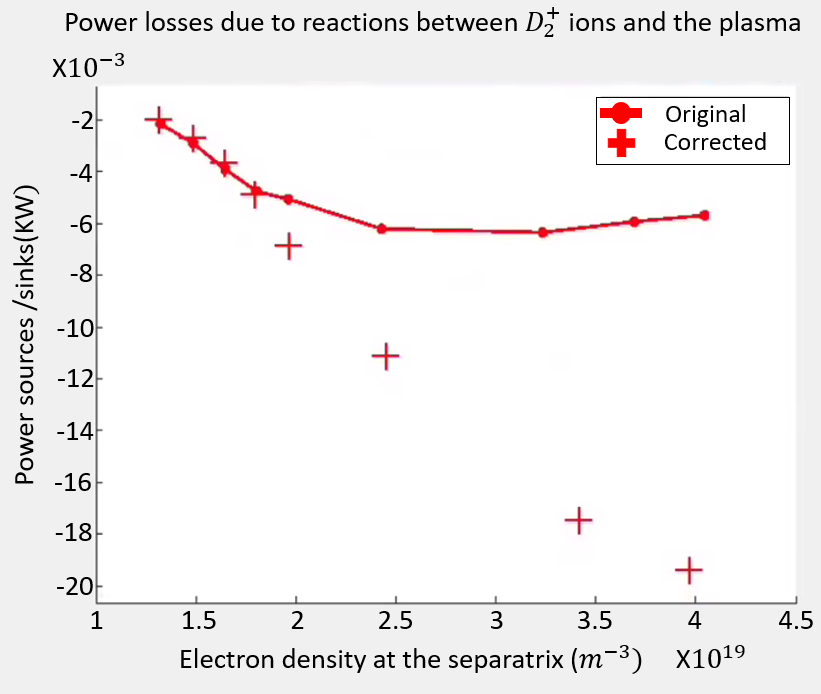}
  \captionof{figure}{The data shows how the interactions between the D2+ ions and the plasma affect the Power balance. We can see a significant increase in the power sink due to the increased density of D2+ density in the corrected rates.}
  \label{fig:Power balance eirene ion plasma el}
\end{figure}

The difference between the post processing and the corrected simulations seen in figure \ref{fig:Post-processed} is known to be due to the lack of self constancy within the post processing routine, and so the difference is due to how increasing the MAR rate has affected the plasma conditions as a whole. Therefore, to understand where these differences originate from we have to observe the changes these effects have on the power, particle and momentum balance of the plasma. By analysing the balance.nc files, we can investigate the changes in the contribution to the power and momentum balance in the plasma. \\

\begin{figure} [!ht]
\centering
  \centering
  \includegraphics[width=.65\linewidth]{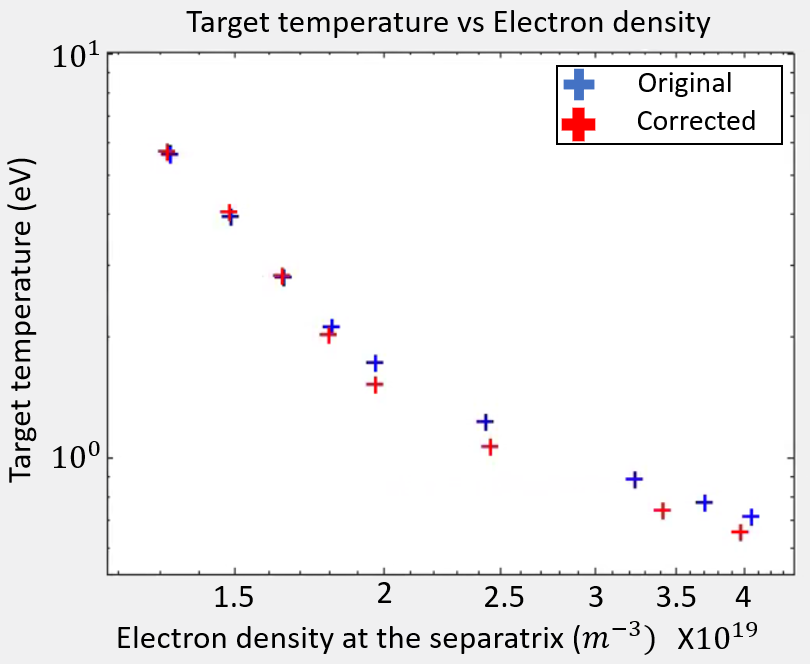}
  \captionof{figure}{Shows how the corrected reaction rate affects the Target temperature compared to the electron density at the seperatrix. We can see a slightly reduced temperature for the corrected rate at higher densities.}
  \label{fig:Target temp vs electron density}
\end{figure}

For the power balance, we can see in figure \ref{fig:Power balance total} there was minimal change in the contributions except for the the reactions between the $D_2^+$ ions and the plasma which are isolated in figure \ref{fig:Power balance eirene ion plasma el}. We can see a significant reduction in the power balance due to $D_2^+$ ions - plasma in the corrected rates at fueling densities which correspond to the significant increases in the molecular charge exchange rate in the corrected rates simulations. This temperature reduction is then confirmed in figure \ref{fig:Target temp vs electron density} where we can see that, for high densities, the corrected rate reduces the target temperature by around 1ev.  From Verhaegh's paper \cite{Main paper on MAR EIR and ionisation rates} we can suggest the majority of this change occurred due to the increase in MAD rate, as MAR leads to minimal cooling compared to MAD created by charge exchange. \\\\

As previously shown in figure \ref{fig:D2+ density ratio}, a lower temperature results in less change exchange, less $D_2^+$ ions and thus less MAR than predicted by the post-processing which doesn't take into account the increased MAR rate reducing the plasma temperature.\\

This also offers an explanation for the reduction in atomic ionisation source in the outer divertor leg compared to both the original simulations and the post-processing, as its reaction rate decreases rapidly with decreasing temperature below 5ev \cite{SOLPS vibrational states for hydrogen}.

However, this is not the only explanation. As detachment occurs the ionisation front moves upstream from the target, it may be that as the ionisation region moves upstream, it moves above the monitored area shown in figure \ref{fig:Tokamak Divertor Grid} and is no longer included in our selection of the divertor gird, and by extension would not be detected by spectrometers observing the divertors. This is more likely to be significant in TCV due to the lack of baffles closing off the divertor. These effects will not be taken into consideration by the post-processing, thus the discrepancy.\\
\begin{figure} [!ht]
\centering
  \centering
  \includegraphics[width=.7\linewidth]{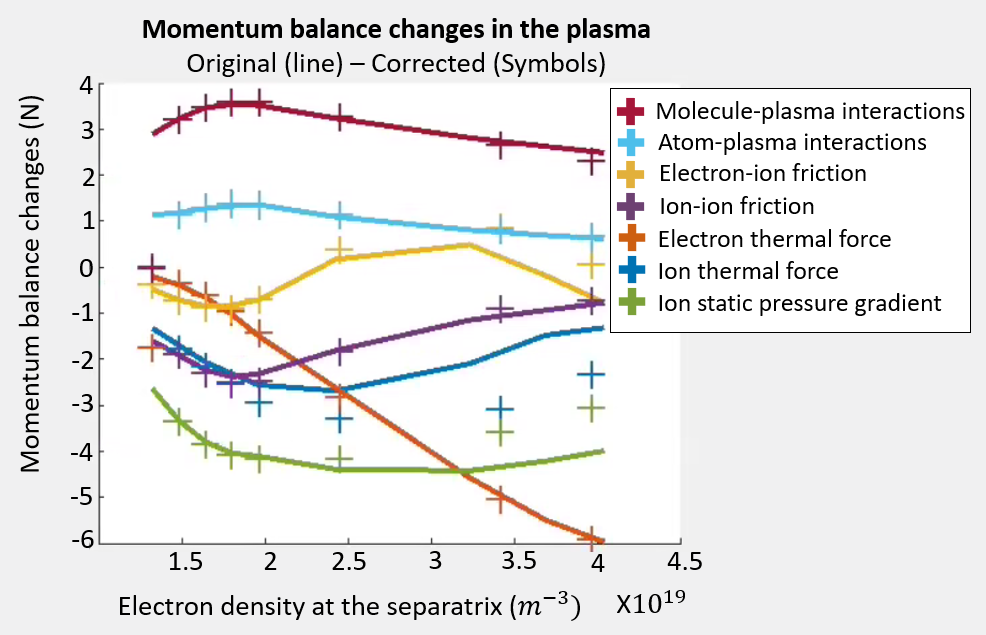}
  \captionof{figure}{The data shows the changes in momentum balance between the corrected and original rates vs the electron density at the seperatrix. We can see minimal changes in the momentum balance between the two. }
  \label{fig:Momentum balance total}
\end{figure}

A significant point is to note that there was minimal change in the momentum balance. In figure \ref{fig:Momentum balance total} we can see that the majority of the contribution to momentum balance remain relatively unchanged by the corrected rates. While there is a minor increase due changes to electron thermal pressure and minor decrease due changes in atom-plasma interactions these two effects also appear to cancel each other out. This results in minimal difference between the original and corrected rates when it comes to the volumetric momentum losses as a function of the target temperature.\\  

\begin{figure} [!ht]
\centering
  \centering
  \includegraphics[width=.65\linewidth]{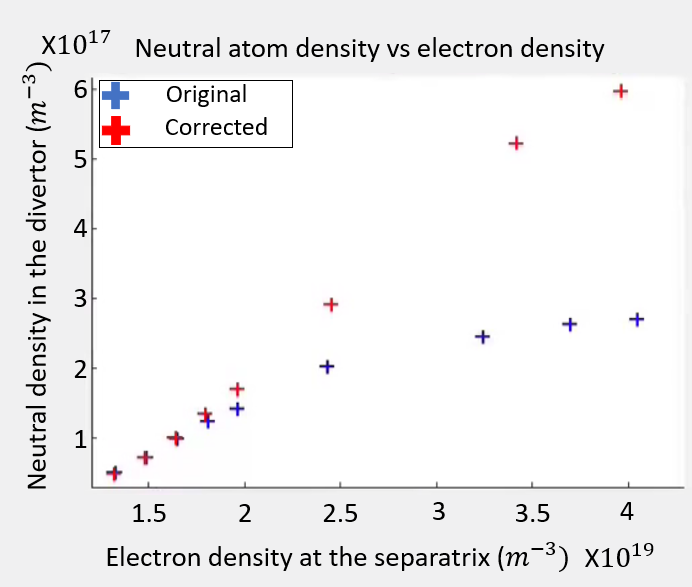}
  \captionof{figure}{We can see the change between the corrected and original rates for the neutral density at the target vs the electron density at the seperatrix. We can see a significantly increased neutral atom density for the corrected rates at high electron densities. }
  \label{fig:MAD}
\end{figure}

A notable observation can be made when investigating the other effects of the increased $D_2^+$ density. We can see in figure \ref{fig:MAD} a significant increase in the neutral density of the plasma relative to the same electron density. This increase is likely due to a significant increase in MAD reactions producing neutral atoms. MAD has a significantly higher reaction rate than MAR \cite{SOLPS vibrational states for hydrogen} and so the corrected rate increasing the $D_2^+$ density results in a strong increase in MAD and thus the neutral atoms it produces.
This is significant because the neutral pressure is sometimes used as a control variable, in the same way that electron density at the seperatrix has been used in this dissertation \cite{Pitt's neutral atom density usage}. The suggestions that simulations have been inaccurately modeling the neutral density may have implications for its use as a reliable control variable in previous investigations.\\\\

Finally, analysing figure \ref{fig:Experimental data}, we can see an agreement in the general process of the plasma that improves significantly from the original simulations. We observe rollover occurring in the particle balance, with both having this occur due most significantly to a power starvation of the ionisation rates and an increase in MAR. We also observe the strong increase in molecular contribution to the $D\alpha$ emission that becomes dominant over the atomic contribution as the fueling density increases. However, we do notice clear differences, most notably is the experimental increase in MAR appear to increase exponentially at higher densities and has an overall larger magnitude. This is likely due to a lacking in the simulations. We can see in Figure \ref{fig:Experimental data} that the experimental data takes into account MAR due to $D^-$ ions which is overlooked in the simulations. If the effects of $D^-$ ions are fully taken into account within simulations, this may close the gap in the MAR reaction rates.

\section{Conclusions}\label{conclusions}

We can see that the correction to the Deuterium reaction rate results in a significant increase in the density of $D_2^+$ ions within the divertor at temperatures of around 1ev. We can also observe that this increase in $D_2^+$ density greatly increases the rate of MAR within the divertor and this induces detachment earlier and deeper in the simulations as a consequence. Further, this increased ion density also affects other reactions, such as the rate of MAD which significantly alters the simulated neutral atom density. Overall, comparing these results to the experimental values we see a far greater agreement compared to the original simulations, as a result we can say these corrected simulations provide a much more accurate prediction of how the plasma will behave in a divertor and when it will detach. This improved ability to accurately predict the behaviour of plasma within the divertor is significant as it may allow for more accurate simulations of how future tokamaks, such as DEMO, will behave and thus influence their design and construction. \\

However, these simulations are not fully accurate compared to the experiment suggesting that further expansions on this experiment are needed. This could include observing how changes in different Tokamak geometries and divertor configurations affects the accuracy of these simulations and it's predicted reaction rate. We could also observe how different wall materials affect the reactions as the plasma particles interact with the wall. Further actives could be the investigations of the effects of including of other plasma factors, such as the role of $D^-$ ions, in simulations. Finally, a full investigation into the reaction rate of molecular charge exchange for deuterium could be performed and even expanded to other reaction rates, due to deuterium's chemical differences from hydrogen. These improved reaction rates could then be implemented into SOLPS-ITER.\\

\section{Appendices}\label{Appendices}

\subsection{Appendix A: Replicability of simulations} \label{Appendix A}
\begin{figure}[!h]
     \centering
     \begin{subfigure}[b]{0.5\textwidth}
         \centering
         \includegraphics[width=.95\linewidth]{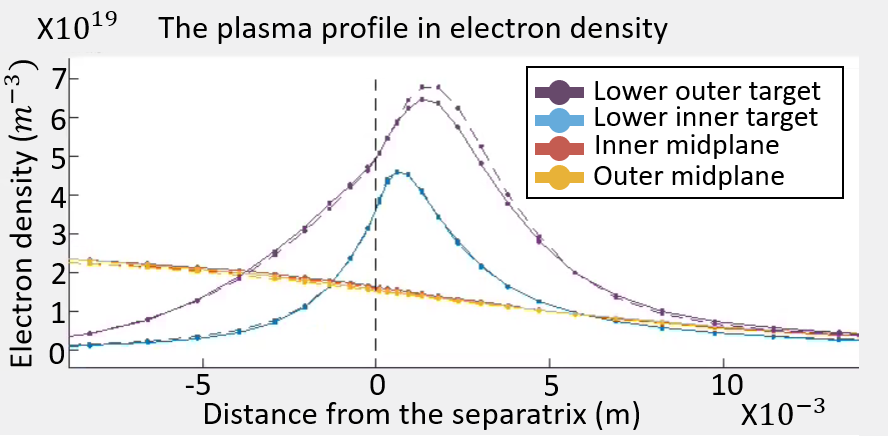}
         \captionof{figure}{}
         \label{fig:e density no current}
     \end{subfigure}
     \hfill
     \begin{subfigure}[b]{0.48\textwidth}
         \centering
         \includegraphics[width=.9\linewidth]{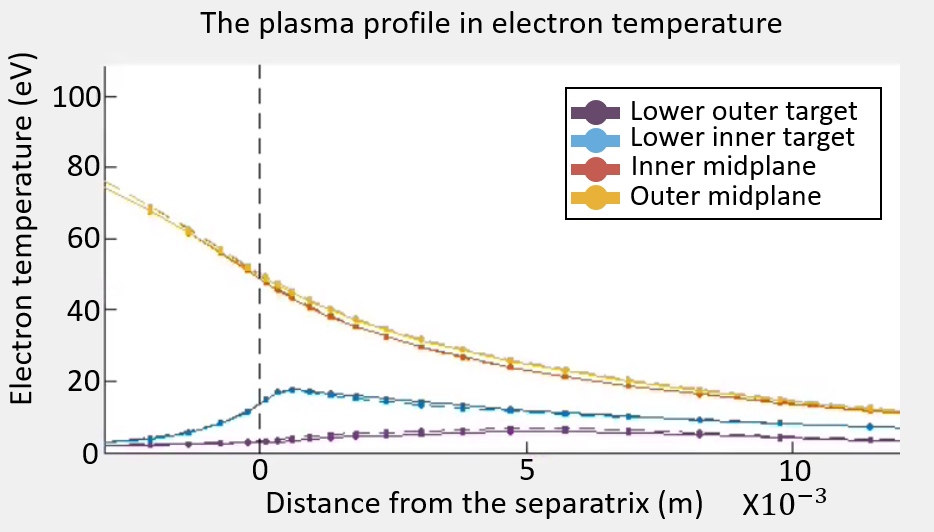}
         \captionof{figure}{}
         \label{fig: temp no current}
     \end{subfigure}
     \hfill
        \caption{A comparison of the plasma profiles for Alex Fil's simulations compared to our modern simulations with the current flow removed with our simulations in block lines and Alex's simulations in dashed. We can see a strong similarity between the two profiles with only a minor difference noticeable in figure \ref{fig:e density no current} in the lower outer targets electron density near the seperatrix}
        \label{fig:No currents}
\end{figure}
The basis of the reliability of these simulations is due to them being continuations of simulations that have been previously published and peer reviewed. However the SOLPS code has changed significantly since the previous simulations in 2017. So we need to validate that these simulations are still similar. The main update to the code has been the introduction of currents to the simulations. To compare the similarity of simulations, we ran our simulations with the original reaction rate and currents removed and compared to Alex's simulations.

When observing the differences we can see there is a strong similarity across the profiles for the key parameters of electron density and the electron temperature. We notice the strongest difference between the two sets of simulations in the electron density of the outer mid-plane but this difference maintains a consistency in the general trends of the plasma and maintains a smaller than 10\% difference in magnitude. Given the various small alterations to the code  over 5 years this minor variance in the simulations was judged to be an acceptable deviation.

\subsection{Appendix B: Hysteresis simulations} \label{Appendix B}
\begin{figure}[!h]
     \centering
     \begin{subfigure}[b]{0.48\textwidth}
         \centering
         \includegraphics[width=.95\linewidth]{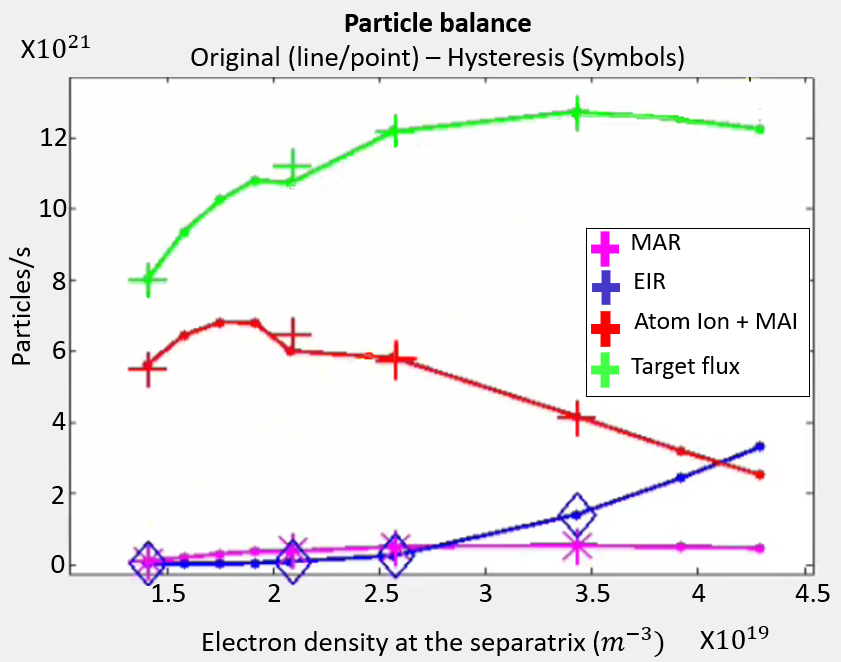}
         \captionof{figure}{}
         \label{fig: Particle balance Hysterises data}
     \end{subfigure}
     \hfill
     \begin{subfigure}[b]{0.5\textwidth}
         \centering
         \includegraphics[width=.95\linewidth]{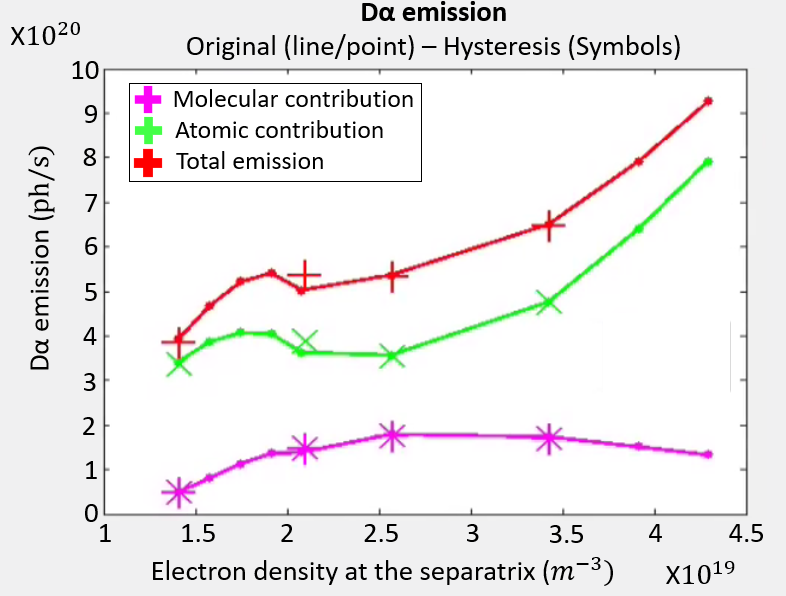}
         \captionof{figure}{}
         \label{fig: Dalpha Hysterises data}
     \end{subfigure}
     \hfill
        \caption{A comparison of the original simulations and the Hysteresis simulations. We can see a minimal difference in the results for particle balance reaction rates and the $D \alpha$ emissions between the original run and the hysteresis runs.}
        \label{fig:Hysterises data}
\end{figure}

The simulations run on a set of differential equations which start at a non-experimental set of initial conditions. These equations are then run for timesteps until the solution reaches a minimum. As a result, it is possible that the our simulations may have converged on a local minimum rather than the global minimum. Therefore a set of hysteresis simulations were created to ensure that, even with altered initial conditions, the simulations converge on the same solution. So the highest fueling density simulation for the original rates (2.4 $\times 10{21}$) was taken and then it's fueling rate altered to the fueling rate a step below. This simulation was then run to convergence and then altered to the next step in fueling rate until all the simulations for the original fueling rates were re-run with new initial conditions.  \\\\

In figures \ref{fig: Particle balance Hysterises data} and \ref{fig: Dalpha Hysterises data} we can see the results of these hysteresis simulations in terms of their Particle balance and $D \alpha$ emissions. What we observe is there there is minimal difference between the original simulations and the hysteresis simulations. This suggests that the simulations have successfully reached a global minimum in their solutions and thus are reliable simulations to use.

\end{document}